\documentstyle[aps,prl,multicol,fancyheadings,epsfig,amsbsy,amssymb,amstex]{revtex}

\def\bbbc{{\mathchoice {\setbox0=\hbox{$\displaystyle\rm C$}\hbox{\hbox
to0pt{\kern0.4\wd0\vrule height0.9\ht0\hss}\box0}} {\setbox0=\hbox{$\textstyle\rm C$}\hbox{\hbox
to0pt{\kern0.4\wd0\vrule height0.9\ht0\hss}\box0}} {\setbox0=\hbox{$\scriptstyle\rm C$}\hbox{\hbox
to0pt{\kern0.4\wd0\vrule height0.9\ht0\hss}\box0}} {\setbox0=\hbox{$\scriptscriptstyle\rm
C$}\hbox{\hbox to0pt{\kern0.4\wd0\vrule height0.9\ht0\hss}\box0}}}} 
\newcommand{\beq}{\begin{eqnarray}}
\newcommand{\eeq}{\end{eqnarray}}

\begin{document}
\title{Measurement induced quantum-classical transition}

\author{D. Mozyrsky and I. Martin}

\address{Theoretical Division, Los Alamos National Laboratory,
Los Alamos, NM 87545, USA}

\date{Printed \today}

\maketitle

\begin{abstract}
A model of  an electrical point contact coupled to a mechanical system
(oscillator) is studied to simulate the dephasing effect of measurement on a
quantum system.  The problem is solved at zero temperature under conditions of
strong non-equilibrium in the measurement apparatus.  For linear coupling
between the oscillator and tunneling electrons, it is found that the oscillator
dynamics becomes damped, with the effective temperature determined by the
voltage drop across the junction. It is demonstrated that both the quantum
heating and the quantum damping of the oscillator manifest themselves in the
current-voltage characteristic of the point contact.
\end{abstract}


\begin{multicols}{2}

There is a dramatic difference in the observed behaviors of microscopic particles and of
macroscopic objects. The everyday-scale objects obey the rules of classical Newtonian mechanics,
while microscopic particles command the use of quantum physics for their description.  The effects
of quantum coherence are almost never observed at the macroscale.  The only known exceptions are
realized when the macroscopic quantum state is particularly robust against external perturbations,
as is the case for superconductors and quantum Hall liquids.  Hence, it appears natural to assume
that it is the coupling to the external world, or $environment$, that leads to $decoherence$ and
consequently to a transition from quantum to classical behavior. This process was explored in
detail in numerous works. It has been shown that within a phenomenological model of environment,
at sufficiently high temperatures, a quantum mechanical system becomes effectively
classical~\cite{fv,CL}. The environment provides both the decoherence and the dissipation needed
for the quantum-classical transition.

Another important distinction between the classical and quantum systems is in their response to
measurement.  Measurement of a classical system in principle can have no effect on the state of
the system; on the other hand, in the quantum regime, the measurement itself is  a source of
decoherence that inevitably changes the state of the system~\cite{book}. The main difference
between the measurement process and the environment induced dephasing is that measurement is an
intrinsically non-equilibrium process.  In this work we demonstrate that despite the apparent
differences, the measurement can also induce a quantum-classical transition.  Recently, Gurvitz
{\em et al.} ~\cite{gurvitz} and Korotkov {\em et al.}~\cite{korotkov} have shown that electrical
measurement leads to dephasing of a finite state system that is being measured.  The systems that
they have studied did not have however a classical analogue. Here, we extend their approach to the
problem of the measurement of a mechanical system. Specifically, we consider a {\it quantum}
oscillator coupled to an electrical point contact in the tunneling regime~\cite{cross}. We solve
this combined problem in the non-equilibrium limit of large voltage across the point contact. From
the general solution, we separately extract the dynamics of the oscillator and of the current
through the contact. For the oscillator, our main findings are that (1) coupling to the tunneling
electrons leads to dissipation with damping coefficient independent of the applied voltage, (2)
current shot noise generates fluctuations and decoherence, and (3) behavior of the mechanical
system becomes effectively classical with the temperature equal to one half of the voltage drop
across point contact. For the point contact, we determine the stationary non-linear
current-voltage ($I$-$V$) characteristic, which implicitly measures the effects of the quantum
heating and quantum dissipation in the oscillator. Thereby we explicitly demonstrate how
measurement can induce a quantum-classical transition.

\begin{figure}[htbp]
\begin{center}
\includegraphics[width = 2.5 in]{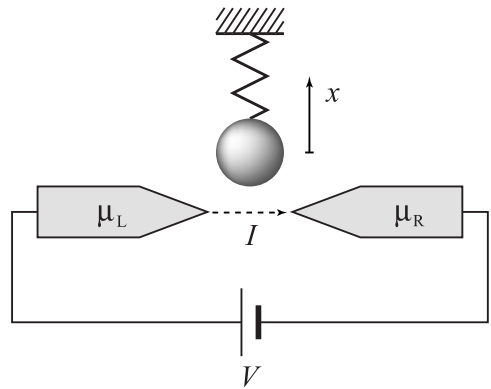}
\vspace{0.5cm} \caption{Model setup.  An oscillator with a confining potential $U(x)$ (shown as
spring) is coupled to an electrical point contact.  The motion of the oscillator modulates the
electron tunneling between the reservoirs.  The carriers in the reservoirs are driven out of
equilibrium by the applied bias voltage, $eV = \mu_L - \mu_R > 0$.}
\label{fig:setup}
\end{center}
\end{figure}

We consider a system described by  the Hamiltonian
\begin{eqnarray}
{\hat H} = \sum_l \epsilon_l a^{\dag}_l a_l &+& \sum_r \epsilon_r a^{\dag}_r a_r + {\hat H}_0(x)\nonumber\\
&+&\sum_{l,r} {\hat \Omega}(x)\left( a^{\dag}_l a_r + a^{\dag}_{r} a_{l} \right) \, . \label{a1}
\end{eqnarray}
Here $a^{\dag}_l$ ($a^{\dag}_r$) creates an electron at the energy level
$\epsilon_l$ ($\epsilon_r$) in the left (right) reservoir.
${\hat H}_0(x) =-(1/2m)\nabla^2_x + U(x)$ is the unperturbed
Hamiltonian of an oscillator of mass $m$ in classical confining potential
$U(x)$, where $x$ is the oscillator coordinate. In what follows we set both
electron charge $e$ and Planck constant $\hbar$ to unity unless stated
otherwise. Generalization to multi-dimensional case when $x$ is a vector is
straightforward. The last term in Eq.~(\ref{a1}) describes tunneling of
electrons between reservoirs, modulated by the oscillator position $x$. For
simplicity we assume that the transition matrix element
${\hat\Omega}(x)$ is independent of single electron states $l$ and
$r$ in the reservoirs.  The latter dependence can be included
without qualitative modification of our principal results. Moreover,
${\hat\Omega}(x)$ can be a general hermitian operator of the particle's
coordinates, i.e. it can be a function of momenta. Finally, there is an
electrical bias,
$V$, applied across the tunnel junction, so that the chemical
potentials in the reservoirs are related as $\mu_L -\mu_R = eV  >
0$.

To determine the time evolution of the system described by the above
Hamiltonian, we use the many-body Shr\"odinger equation approach developed in
Refs.~\cite{gurvitz} for solution of highly-nonequilibrium quantum transport
problems.  The wavefunction of the full system is
\begin{eqnarray}
|{\bf \Psi}(t)\rangle = \Big\{&& b_0(x,t) + \sum_{l,r} b_{lr}(x,t) a^{\dag}_r a_{l}\nonumber\\
& & + \sum_{l,l^{\prime}r,r^{\prime}} b_{ll^{\prime}rr^{\prime}}(x,t) a^{\dag}_{r}
a^{\dag}_{r^{\prime}}a_{l}a_{l^{\prime}} + \  \ldots \ \Big\}|{\bf 0} \rangle\, , \label{a2}
\end{eqnarray}
where the initial state $|{\bf 0} \rangle$ corresponds to fully occupied Fermi seas in the
reservoirs, with no electrons above the respective chemical potentials.  This wavefunction is a
superposition of all possible electron-hole combinations that can be generated by the Hamiltonian
$H$; note that $H$ conserves the total number of electrons in the reservoirs.  The first term in
${\bf\Psi}$ is the time-dependent part of the wavefunction that corresponds to unchanged occupancy
of Fermi seas; the second term describes a state in which a hole is created in the left reservoir
and an extra electron occupies the right reservoir, etc. In our representation, the amplitudes
$b(x,t)$ explicitly depend on the coordinate $x$ of the oscillator. Substituting
wavefunction~(\ref{a2}) into the Scr\"odinger equation with Hamiltonian~(\ref{a1}), $i|{\bf \dot
\Psi} \rangle = H |{\bf \Psi} \rangle$, and performing imaginary Laplace transform on ${\bf\Psi}$,
${\tilde b}(\omega)= \int_0^\infty dt e^{i\omega t}b(t)$, we obtain an infinite set of equations
for the amplitudes ${\tilde b}$:
\begin{mathletters}
\label{a3}
\begin{eqnarray}
&&\left(\omega -{\hat H}_0(x)\right) {\tilde b}_0(x,\omega) - \sum_{l,r}
{\hat \Omega}(x) {\tilde b}_{lr}(x,\omega)=i \psi_0(x)\, ,\label{a3a}\\
&&\left(\omega -{\hat H}_0(x) +\epsilon_l-\epsilon_r\right){\tilde
b}_{lr}(x,\omega) - {\hat \Omega}(x){\tilde b}_0(x,\omega)\nonumber\\
& & \quad\quad\quad\quad\quad\quad\quad\quad\quad\quad - \sum_{l^{\prime},r^{\prime}} {\hat
\Omega}(x)
{\tilde b}_{ll^{\prime}rr^{\prime}} (x,\omega) = 0\, , \label{a3b}\\
& & \quad\quad\quad\quad\quad\quad\quad\quad\quad\quad\quad \ldots \quad .\nonumber
\end{eqnarray}
\end{mathletters}
Here, $\psi_0(x)$ is the initial state of the particle.  In order to proceed, we solve the
$n^{th}$ equation for the $n^{th}$ amplitude ${\tilde b}(\omega, x)$ in terms of the $(n-1)^{th}$
and $(n+1)^{th}$ amplitudes and substitute it into the $(n-1)^{th}$ equation.  Then, as it was
shown in Ref.~\cite{gurvitz}, the contribution of the $(n+1)^{th}$ amplitude to the $(n-1)^{th}$
equation is negligible. Not only it is of higher order in the coupling constant ${\hat\Omega}(x)$,
but it also vanishes in the limit of large bias.  As an example, let us solve Eq.~(\ref{a3b}) for
${\tilde b}_{lr}(\omega, x)$ and substitute the resulting expression into Eq.~(\ref{a3a}). The sum
in Eq.~(\ref{a3a}) becomes proportional to ${\tilde b}_0$,
\begin{eqnarray}
\sum_{l,r,n}\int dx^{\prime} {{\hat \Omega} (x) {\hat \Omega}
(x^{\prime})\phi_n(x)\phi_n^{\ast}(x^{\prime})\over \omega - E_n+\epsilon_l-\epsilon_r}{\tilde
b}_0(\omega, x^{\prime})\ ,\label{a30}
\end{eqnarray}
where $\phi_n(x)$ and $E_n$ are eigenfunctions and eigenvalues of ${\hat H}_0(x)$, ${\hat H}_0
\phi_n(x)=E_n \phi_n(x)$. Replacing summations over $l$ and $r$ by integrals over the corresponding
densities of states, $\rho_L$ and $\rho_R$ (here assumed constant), $\sum_{l,r}(\omega -
E_n+\epsilon_l-\epsilon_r)^{-1} = -i\pi\rho_L\rho_R(V + \omega - E_n) \Theta (V + \omega - E_n)$.
The real part of the sum only weakly renormalizes the self energy of the
oscillator~\cite{realpart}. The step-function $\Theta$ can be dropped provided the typical energy
spacing for ${\hat H}_0$ is much smaller than $V$ and the particle's dynamics involves only
relatively low-lying energy levels, such that condition $V > E_n$ is satisfied. The latter
assumption is consistent with the final results of our calculation and thus $\Theta (V + \omega -
E_n)$ is omitted. Then, using the completeness relation for $\phi_n(x)$, the above sum yields
$-i\pi\rho_L\rho_R [{\hat \Omega}^2(x)(V + \omega) - {\hat \Omega}(x){\hat H}_0(x){\hat
\Omega}(x)]{\tilde b}_0(\omega, x)$. Applying this procedure to all equations in the chain
~(\ref{a3}) and neglecting terms of order $O({\hat\Omega}^4)$ and higher, we obtain a set of
simplified equations,
\begin{mathletters}
\label{a4}
\begin{eqnarray}
&&\left(\omega -{\hat H}_0(x)\right) {\tilde b}_0(x)\nonumber\\
&&\quad\quad\quad-i{\eta \over 2}\Big\{{\hat \Omega}\left[{\hat H}_0, {\hat \Omega} \right](x)
-V{\hat \Omega}^2(x)\Big\}{\tilde b}_0(x) = \Delta(x)\, ,\label{a4a}\\
&&\left(\omega -{\hat H}_0(x) +\epsilon_l-\epsilon_r\right){\tilde b}_{lr}(x) - {\hat \Omega}(x)
{\tilde b}_0(x)\nonumber\\
&&\quad\quad\quad-i{\eta \over 2} \Big\{{\hat \Omega}\left[{\hat H}_0, {\hat \Omega}\right](x)
-V{\hat \Omega}^2(x)\Big\}{\tilde b}_{lr}(x) = 0\, ,\label{a4b}\\
&&\quad\quad\quad\quad\quad\quad\quad\quad\quad\quad\quad \ldots \quad ,\nonumber
\end{eqnarray}
\end{mathletters}
where square brackets denote commutator, ${\hat \Omega}[{\hat H}_0, {\hat
\Omega}](x) \equiv {\hat
\Omega}(x)[{\hat H}_0(x), {\hat \Omega}(x)]$, $\eta = 2\pi \rho_L\rho_R$, and $\Delta (x)
=i\psi_0(x)(1-i\pi\rho_L\rho_R{\hat \Omega}^2(x))$.
Eqs.~(\ref{a4}) allow us to derive the quantum rate equations for
our system. We introduce the density matrix of the system defined
as
\beq
\sigma^0(x,x^{\prime},t) &=&
b_0(x,t)b^{\ast}_0(x^{\prime},t), \nonumber\\
\sigma^1(x,x^{\prime},t) &=& \sum_{l,r} b_{lr}(x,t)b^{\ast}_{lr}(x^{\prime},t),\nonumber\\
&&\ldots \quad .\nonumber
\eeq
Each of these objects is the density matrix of
the oscillator conditioned by the number of electrons (the superscript in
$\sigma$'s) that have arrived  in the right reservoir. Here we outline the
derivation of the rate equation for
$\sigma^1$. After replacing $x$ and $\omega$ in Eq.~(\ref{a4a}) by
$x^{\prime}$ and $\omega^{\prime}$ and multiplying this equation
by ${\tilde b}_{lr}(\omega, x)$, we subtract the result from
Eqs.~(\ref{a4b}) multiplied by ${\tilde
b}^{\ast}_{lr}(\omega^{\prime}, x^{\prime})$. Summing over
$l$ and $r$ and applying the inverse Laplace transform to the resulting equation, we obtain
\begin{eqnarray}
{d \over dt}&& \sigma^1= -i\left[{\hat H}_0,\sigma^1\right]+{\eta
\over 2}\Big\{{\hat \Omega} \left[{\hat H}_0, {\hat \Omega}\right](x) - V{\hat \Omega}^2(x)\Big\}\sigma^1~~\nonumber\\
&& +{\eta \over 2}\Big\{ {\hat \Omega}\left[{\hat H}_0, {\hat \Omega}\right](x^{\prime}) - V{\hat
\Omega}^2(x^{\prime})\Big\}\sigma^1 +2{\rm Im}\left({\hat \Omega}(x)\sigma^{0,1}\right),\nonumber
\end{eqnarray}
where we introduced the off-diagonal element $\sigma^{0,1}(x,x^{\prime},t)=
\sum_{l,r}b_0(x,t)b_{lr}^{\ast}(x^{\prime},t)$. The latter can be expressed in terms of the
diagonal element $\sigma^0$. Following the above described procedure, we find
\[
\sigma^{0,1} = i{\eta \over 2}\Big\{V{\hat \Omega}(x^{\prime}) - \left[{\hat H}_0, {\hat \Omega}
\right](x^{\prime})\Big\}\sigma^0\ .
\]
Substituting this expression into the equation for $\sigma^1$ we obtain the rate equation that
couples $\sigma^1$ and $\sigma^0$ density matrices. The result can be generalized to the density
matrix $\sigma^n$ for any number $n$ of electrons that tunneled across the point contact,
\begin{eqnarray}
{\dot \sigma}^n =&-&i\left[{\hat H}_0,\sigma^n\right]+ {\eta \over 2}\Big\{{\hat \Omega}
\left[{\hat H}_0, {\hat \Omega}\right](x)+{\hat \Omega} \left[{\hat H}_0,
{\hat \Omega}\right](x^{\prime})\nonumber\\
&-& V\left({\hat \Omega}^2(x)+{\hat \Omega}^2(x^{\prime})\right)\Big\}\sigma^n -{\eta \over
2}\Big\{{\hat \Omega}(x)
\left[{\hat H}_0, {\hat \Omega}\right](x^{\prime})\nonumber\\
&+& {\hat \Omega}(x^{\prime})\left[{\hat H}_0, {\hat \Omega}\right](x)-2V{\hat \Omega}(x){\hat
\Omega}(x^{\prime})\Big\}\sigma^{n-1}\ .\label{a5}
\end{eqnarray}

From the equations of motion Eqs.~(\ref{a5}) for the full density matrix, one can obtain a closed
form equation for the density matrix of the oscillator alone, $\sigma(x,x^{\prime},t)$. Summing
Eqs.~(\ref{a5}) over $n$ we find that the equation for $\sigma$ can be written in an invariant
form,
\begin{eqnarray}
{\dot \sigma} =-i\left[{\hat H}_0,\sigma\right] + {\eta \over 2}\left[{\hat \Omega},\Big\{{\hat
\Lambda}, \sigma \Big\}\right] - {V\eta \over 2}\left[{\hat \Omega},\left[{\hat
\Omega},\sigma\right]\right]\ ,\label{a6}
\end{eqnarray}
where ${\hat \Lambda}=[{\hat H}_0, {\hat \Omega}]$ and curly brackets denote anticommutator.
Similarly, one can determine the time dependence of the average current in terms of the oscillator
density matrix. The current is defined as $\langle I(t) \rangle = \langle {\dot N}(t) \rangle$,
where $\langle N(t) \rangle = \sum_n n\int dx \sigma^n (x,x,t)$ is the expectation value of the
number of electrons that have tunnelled into the right contact by time $t$. Using Eqs.~(\ref{a5})
for $\dot{\sigma}^n$, we find the current
\begin{eqnarray}
\langle I(t) \rangle = V\eta{\rm Tr}\left({\hat \Omega}^2\sigma(t)\right) - {\eta \over 2}{\rm
Tr}\left( \left[{\hat \Omega},{\hat \Lambda}\right] \sigma(t)\right),\label{a7}
\end{eqnarray}
in terms of the oscillator density matrix.  Eqs.~(\ref{a6}) and~(\ref{a7}) constitute the
principal results of this work. They describe the evolution of a system modified by its interaction
with non-equilibrium environment, as well as influence of the oscillator on the current between
reservoirs.  Remarkably, for linear coupling between the two subsystems, $\Omega(x) = \Omega_0 +
Cx$, we recover the Caldeira and Leggett equation (CL)~\cite{CL} for the density matrix of the
oscillator. Indeed, for the linear coupling, the second term in the RHS of Eq.~(\ref{a6}) produces
the dissipative term $(\eta C^2/2m)(x-x^{\prime})(\partial_{x^{\prime}}-\partial_x)\sigma$, while
the last term becomes $(\eta V C^2/2)(x-x^{\prime})^2\sigma$. It is responsible for
fluctuation/decoherence induced by the tunnel current.

Both the CL equation and our Eq.~(\ref{a6}), describe high temperature dynamics of a mechanical
system interacting with a heat bath. There is a substantial difference, however.  Unlike the CL
work where the heat reservoir is in equilibrium at high temperature $T$, our calculation is at T =
0. The effective temperature that arises in Eq.~(\ref{a6}) is not the temperature of the
reservoirs, but rather is a result of the {\em non-equilibrium} fluctuations that arise in the
course of the evolution of the full system described by Hamiltonian~(\ref{a1}). In this
``classical'' limit, the oscillator dynamics specified by the master equation~(\ref{a6}), is given
by the Langevin equation, $ m{\ddot x} + \gamma m {\dot x} + \nabla_x U(x) = f(t)$, where $f(t)$ is
white noise satisfying $\langle f(t) f(t^{\prime})\rangle = 2m\gamma T_{\rm
eff}\delta(t-t^{\prime})$. Comparing Eq.~(\ref{a6}) for ${\hat \Omega} = \Omega_0 + Cx$ with
Caldeira and Leggett equation~\cite{CL}, we deduce the damping coefficient
$$\gamma=\eta \hbar C^2/2m$$
and the effective temperature
$$T_{\rm eff}=eV/2.$$
The fluctuations that give rise to the effective temperature can be traced back to the shot noise
of the tunnel current.

In a real physical system,  one can always expect presence of additional sources of dissipation.
If the current across the junction is the dominant source of the fluctuations for the oscillator,
the effective temperature will remain proportional to the voltage across the junction, $T^\ast =
T_{\rm eff}\gamma/(\gamma +\gamma^{\prime})$, where $\gamma^{\prime}$ is the extra damping in the
system.   In this regime, the measurement of the reduced effective temperature combined with a
separate determination of the total damping coefficient, can provide the value of the
non-equilibrium {\em measurement-induced} dissipation $\gamma$.

Due to the self-consistent coupling between the oscillator and the tunnel current, the
measurement-induced effective oscillator temperature can be directly extracted from the non-linear
part of the $I$-$V$ characteristic. The current dynamics is given by Eq.~(\ref{a7}). The first
term on the RHS of Eq.~(\ref{a7}) is simply the Landauer formula with the transmission coefficient
$\eta\Omega^2(x)$ modulated by the position of the oscillator. This is what one would expect from
the Hamiltonian~(\ref{a1}) assuming elastic electron tunneling.  The second term, however, is
non-trivial and accounts for the possible inelasticity of electron tunneling due to the coupling
to the oscillator.  For linear coupling $\hat{\Omega}(x) = \Omega_0 + Cx$, this term yields
$(-\gamma)$. Hence, the expression for the time-dependent current is
\begin{eqnarray}
\langle I(t) \rangle = {e^2\over\hbar} \eta V(\Omega_0^2 + C^2\langle x^2(t)\rangle)
 -e\gamma.\label{a10}
\end{eqnarray}

For a linear oscillator with frequency $\omega_0$, such that $U(x) =
(m\omega_0^2/2)x^2$, the stationary $I$-$V$ characteristic can be determined by
solving Eq.~(\ref{a6}) for the stationary state.  In the absence of any
additional sources of dissipation, we obtain that in the stationary regime
$\langle x^2\rangle = T_{\rm eff}/(m\omega_0^2)$.  This generates a quadratic in $V$ term in
the $I$-$V$ characteristic, which is a signature of the ``quantum heating'' of
the oscillator.  The constant negative offset term $(-e\gamma)$ is more subtle
and has to be properly interpreted.  Indeed, for low bias this term could
dominate, which would seem to imply a current flowing in the direction opposite
to the applied voltage. This is however only an artifact which signifies the
break down of our approach at low biases.  To trace the origin of this term we
calculate the tunnel current in the low voltage limit at zero temperature from
the linear response theory. Assuming that the oscillator is in the ground
state, we obtain
\[
\langle I\rangle_{\rm L.R.} = {e^2\over\hbar}\eta \Omega_0^2 V
+ {\eta e C^2\over\hbar}\langle x^2\rangle_0(eV - \hbar \omega_0)\Theta(eV - \hbar \omega_0),
\]
where $\langle x^2\rangle_0$ is the average square of the zero point motion.  The
$\Theta$-function indicates that at zero temperature, for small bias neither
the oscillator is excited by the current, nor the current is affected by the
oscillator's zero point motion.  For a bias exceeding the oscillator excitation
energy, an additional channel in the tunneling opens up. This new channel
however requires the excitation of the oscillator, which introduces
inelasticity and effectively reduces the applied voltage by $\hbar\omega_0$.
Since the negative offset term in the linear response equation is identical to
the
$(-e\gamma)$ term that appears in Eq.~(\ref{a10}), we conclude that
both can be attributed to the dissipative effect of the oscillator on the
tunnel current.

Given the correspondence between the large-voltage expression and the linear response result, we
expect that the expression for current, Eq.~(\ref{a10}), remains approximately valid for arbitrary
voltage down to $eV = \hbar\omega_0$. Below this threshold voltage, oscillator and point contact
become decoupled and the traditional expression for the tunnel current in a point contact applies,
assuming that the oscillator is initially in a ground state. In the presence of additional
dissipation, $\gamma^{\prime}$, the square displacement expectation value should be computed at
the reduced effective temperature, $\langle x^2\rangle = eV\gamma/[2(\gamma
+\gamma^{\prime})m\omega_0^2]$. With this modification included, the current-voltage
equation~(\ref{a10}) could be used to fit experimental data in order to extract the
measurement-induced dissipation coefficient $\gamma$.

Based on the above discussion, it would seem impossible to measure zero point
fluctuations of the oscillator.  At low voltages,
$eV <\hbar \omega_0$, while the oscillator is in the ground state, it is completely
decoupled from current.  On the other hand, at high voltages the oscillator is
no longer in the ground state.  We will now demonstrate that it is indeed
possible to measure the effect of zero point fluctuations using {\em
time-dependent} measurement of the $I$-$V$ characteristic.  Assuming that at
$t=0$ oscillator was in its ground state and solving the time-dependent density
matrix equation (\ref{a6}) for a linear oscillator, we find that the
time-dependent oscillator dispersion is
\begin{eqnarray}
\langle x^2(t)\rangle = \langle x^2\rangle_0 Q(t) +\langle x_{cl}^2(t)\rangle,\nonumber
\end{eqnarray}
where $\langle x^2\rangle_0$ is the oscillator ground state dispersion and $Q(t) = e^{-\gamma
t}(\sin^2(\omega_0 t +\phi_0) + \sin^2\omega_0 t)$, with $\cos\phi_0=\gamma/2\omega_0$.  The term
$\langle x_{cl}^2(t)\rangle$ arises due to the ``quantum heating.''  It coincides with the
solution of the classical Langevin equation with the damping coefficient $\eta$ and temperature
$T_{\rm eff}$ for initial conditions $x(0)=0$ and ${\dot x}(0)=0$, $\langle x_{cl}^2(t)\rangle =
(\gamma eV/m\omega_0^2)\int_0^t d\tau e^{-\gamma\tau} \sin^2\omega_0\tau$ for $\omega_0 \gg
\gamma$. In the stationary state for $t=\infty$, we obtain results discussed above.  For small
times, $t\ll\gamma^{-1}$, the $\langle x_{cl}^2(t)\rangle$ term can be neglected and thus from
Eq.~(\ref{a10}) we obtain
\begin{eqnarray}
\langle I(t \ll \gamma^{-1}) \rangle = {e^2\over\hbar} \eta V(\Omega_0^2 + C^2\langle x^2\rangle_0)
 -e\gamma.\
\end{eqnarray}
Hence we find that the zero-point motion can be extracted as the change in the
linear part of the $I$-$V$ characteristic from short to long time dynamics.

A setup similar to the one considered here, was recently proposed in Ref.~\cite{schwab}, where the
authors argued that single electron transistor (SET) can be used as a displacement detectors for a
charged mechanical cantilever.  It was suggested that such setup could provide the resolution
sufficient to detect the cantilever zero point motion.  We expect that despite the differences in
the proposed systems, our framework and results will be relevant for the SET configuration as
well.  Another system that may prove to be suitable for observation of the effects discussed here
is a molecule trapped near a point contact~\cite{mceuen}.

To summarize, we have proposed a model for the coupled mechanical system and
electrical point contact to simulate a measurement-induced quantum-classical
transition.  We have solved the problem under the conditions of a strong
non-equilibrium in the measurement apparatus.  Our main findings are that for
linear coupling between the mechanical oscillator and the tunneling electrons,
the dynamics of the oscillator becomes damped, with the effective temperature
determined by the voltage drop across the junction.  We have also demonstrated
how both the quantum heating and the quantum damping of the oscillator manifest
themselves in the current-voltage characteristic of the point contact.

We thank Keith Schwab for provoking our interest in quantum-electro-mechanical systems. We also
benefited from discussions with G. P. Berman, S. Chakravarty, S. A. Gurvitz, S. Habib and A.
Shnirman. This work was supported by the US DOE. D.M. was also supported, in part, by the US NSA
and the US NSF.

\end{multicols}

\begin{references}
\bibitem{realpart} The real part of the sum~(\ref{a30}) introduces a negative offset term
$-(\ln2\,/W)\Omega^2(x)$ in the self-Hamiltonian of the oscillator. Here $W$ is the bandwidth of
the electronic subsystem, $W\gg V$. While insignificant for a linear oscillator, this term may be
important for anharmonic potentials $U(x)$, as it can strongly affect tunneling processes.
\end{references}
\end{document}